# A three-phase framework for photon scattering in porous nanocomposites

Khalid Alhammadi[a,b,‡], Daniel Carne[c,‡,*], Xiulin Ruan[a,*]

[a] School of Mechanical Engineering, Purdue University, West Lafayette, IN 47907, USA
[b] Mechanical Engineering Department, King Saud University, P.O. Box 800, Riyadh, 11421, Saudi Arabia
[c] Department of Mechanical and Nuclear Engineering, United States Naval Academy, Annapolis, MD 21402, USA
[‡] These authors contributed equally to this work.

[*] Corresponding Authors: ruan@purdue.edu, carne@usna.edu

ABSTRACT

Light scattering in heterogeneous media underpins a broad range of natural phenomena and optical technologies, yet remains difficult to understand in porous nanocomposites. State-of-the-art theories generally treat these materials as two-phase systems consisting of particles and a matrix, overlooking the air pores that are intrinsically present in porous composites. Here we experimentally determine the scattering coefficients of six $BaSO_4$-acrylic and $TiO_2$-acrylic coatings spanning a range of particle volume fractions and develop a three-phase scattering framework that explicitly accounts for pore-induced scattering. By representing the composite as two coupled scattering subsystems, particle-matrix and particle-air, the framework resolves discrepancies between theory and experiment: conventional two-phase models substantially underpredict scattering in $BaSO_4$ coatings while overpredicting scattering in $TiO_2$ coatings, whereas the three-phase model accurately captures the observed trends and reduces the mean absolute logarithmic error from 0.667 to 0.379. This framework identifies air pores as a fundamental and independently controllable contributor to optical scattering, providing new physical insights and design principles for engineering highly scattering porous nanocomposites.

## INTRODUCTION

Accurate modeling of spectral radiative properties is a critical step toward optimizing nanocomposite paint and coating systems, which are central to passive radiative cooling [1], [2] and photovoltaic energy conversion [3]. A nanocomposite medium comprises of at least two optically distinct phases: a continuous matrix and one or more dispersed particulate or pore phases. The structural and optical parameters of these phases set the local scattering coefficient, absorption coefficient, and phase function that enter the radiative transfer equation (RTE) governing the macroscopic response [4], [5], [6]. Two-phase optical models have been extensively developed and validated across metallic and dielectric systems [7], [8], [9]. These models estimate the radiative properties for a representative volume element of randomly distributed particles from the pigment–matrix relative refractive index and the particle size parameter. The interparticle-clearance-to-wavelength ratio then determines whether the independent or dependent scattering regime applies. Yet all share one assumption: the particles are embedded in a single continues binder, so a single particle-matrix interface governs the scattering.

This assumption breaks down in high-solid-volume-fraction radiative cooling paints, where air pores emerge and produce a three-phase nanocomposite of pigment, binder, and air voids. These phases form a complex, interwoven morphology, as shown in the SEM images of Fig. 1. A conventional commercial paint contains only pigment and matrix (Fig. 1a), whereas the high-loading radiative cooling paint develops a third air-pore phase (Fig. 1b). The large index contrast between air (n ≈ 1) and the surrounding pigment and matrix creates scattering interfaces absent from two-phase cases, so that the photon mean free path is no longer set by a single pigment–matrix interface but by the collective geometry and contrast of all three phases.

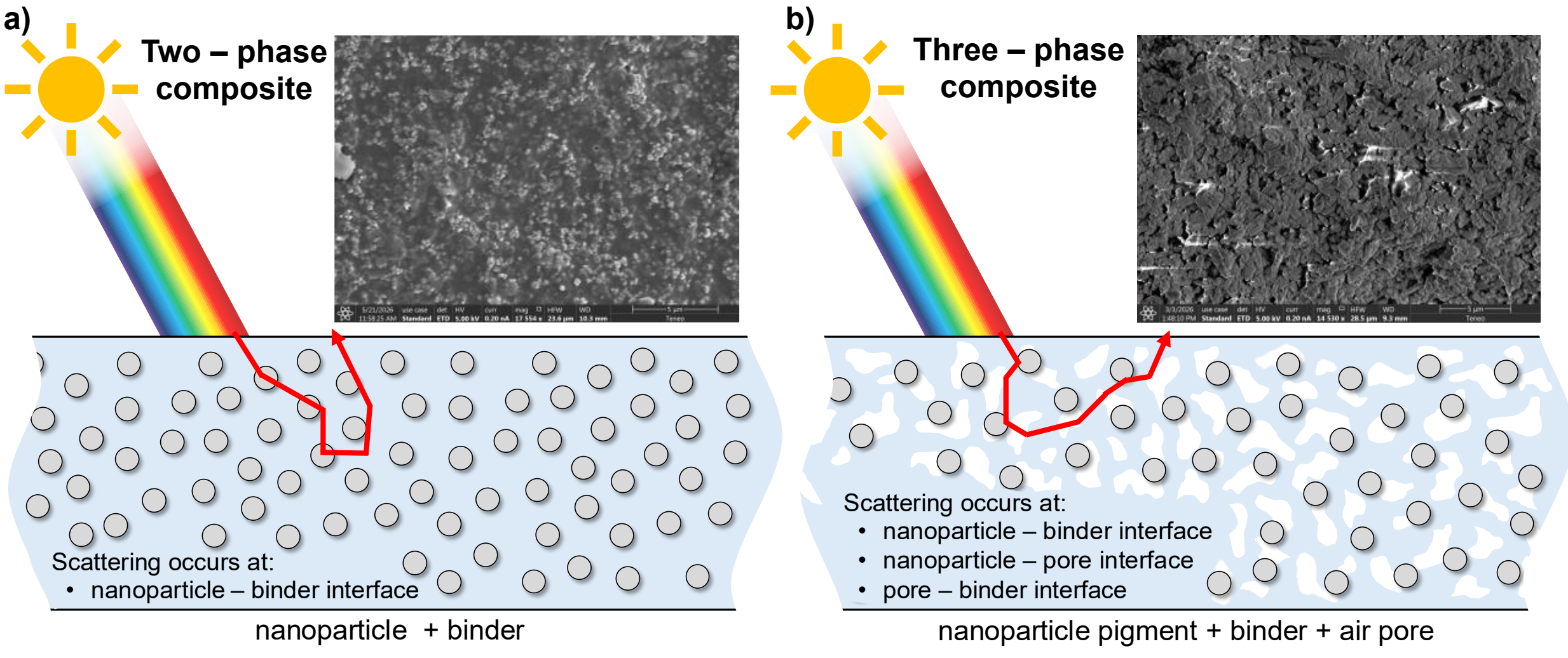

Figure 1: Schematic illustration and SEM images of (a) commercial paint represented as a two-phase system and (b) $BaSO_4$ paint represented as a three-phase system.

The conventional approach to capturing air-pore effects retains a two-phase pigment–matrix model and applies an augmented correction that raises the scattering coefficient to account for the air pores [1], [2], [10]. This correction, however, is a function of the filler volume fraction alone and contains no explicit term for the porosity. In the actual nanocomposite medium, the strength of the air-pore contribution to optical performance depends on the refractive-index contrast between pigment, binder, and air. This limitation also raises the possibility that apparent agreement of two-phase models with experiment could reflect coincidental cancellation between air-pore and dependent-scattering effects rather than true physics. A model that explicitly represents the pigment, binder, and air phases and their optical contrast is therefore needed [1], [10], [11].

Capturing this behavior requires representing the air phase through its coupling with pigment and binder rather than as discrete inclusion. This is because the interconnected morphology of the air phase is difficult to characterize as discrete shapes, and treating the pores as spheres substantially underpredicts the scattering coefficient relative to experiments. However, such a rigorous three-phase optical model for paint systems remains largely absent from the literature. This represents a significant gap for the design of high-performance coatings, in which porosity is either intentionally engineered or an inherent consequence of the nanoparticle composite formulations.

In this study, we experimentally characterize six porous nanoparticle coatings and propose a three-phase model to evaluate the role of air pores in nanocomposite scattering media. $BaSO_4$-acrylic and $TiO_2$-acrylic coatings are fabricated at varying pigment volume fractions and optically characterized from 0.4 – 2.5 μm wavelength. The measured spectral response is then used to derive the reduced scattering and absorption coefficient through an inverse Monte Carlo method, enabling direct comparison between experimentally observed scattering and model predictions. These results are compared against existing two-phase models and the proposed three-phase model. While the two-phase models cannot consistently replicate the trends seen experimentally, the three-phase model accurately captures that air pores significantly enhance scattering in the $BaSO_4$-acrylic coating but not in the $TiO_2$-acrylic coating. We also note that while this three-phase model improves accuracy over the two-phase model and explains trends seen experimentally, it remains a simplified representation of the complex three-phase nanocomposite medium. Accordingly, these experimental measurements and model comparisons are anticipated to provide a foundation for further development of more rigorous three-phase models and dependent scattering corrections for porous nanoparticle composite coatings.

## RESULTS

We now quantify the microscopic radiative properties of the coatings and compare the model predictions with experiment. The hard dielectric pigment particles are treated as the dispersed scattering phase, while the surrounding matrix and air form an effective medium. This choice is motivated by the hard pigment particles keeping their spherical shape on drying, whereas the matrix and air develop a complex, interwoven morphology (SEM images in supplementary figure 2, 3). Scattering at the binder–air interfaces is neglected due to their complex topology. The resulting microscopic scattering and absorption coefficients and asymmetry parameter are propagated through Monte Carlo simulation to obtain the spectral reflectance, transmittance, and absorptance, which we compare against experiment below.

The existing two-phase and the newly proposed three-phase models are compared in Fig. 2, which shows the governing equations for each. The conventional two-phase model (Fig. 2a) accounts only for the solid volume fraction of pigment in a single host. The three-phase model (Fig. 2b) instead partitions the pigment particles between the two surrounding phases: each particle is assumed to be embedded in either the matrix or an air pore, with the fraction in each set by the matrix and air-pore volume fractions ($R_m$ and $R_a$). The scattering efficiency, absorption efficiency, and asymmetry parameter are computed separately for the air-embedded and binder-embedded populations using absorbing-medium Mie theory [12], then volume-averaged by R_m and R_a into a single set of bulk coefficients, with an additional term for matrix absorption. This is a post-Mie effective-medium treatment: the two Mie solutions are combined after the calculation, rather than averaging the air and matrix refractive indices beforehand. The pre-Mie alternative fails when the particle index lies between those of air and binder, which motivates the post-Mie choice. These bulk coefficients are propagated through the Monte Carlo solver in FOS to obtain the spectral response [13]. Because dependent scattering in three-phase media is complex, the model assumes independent scattering; its aim is to isolate and explain the air-pore effects observed in experiment, with dependent-scattering treatments left to future work.

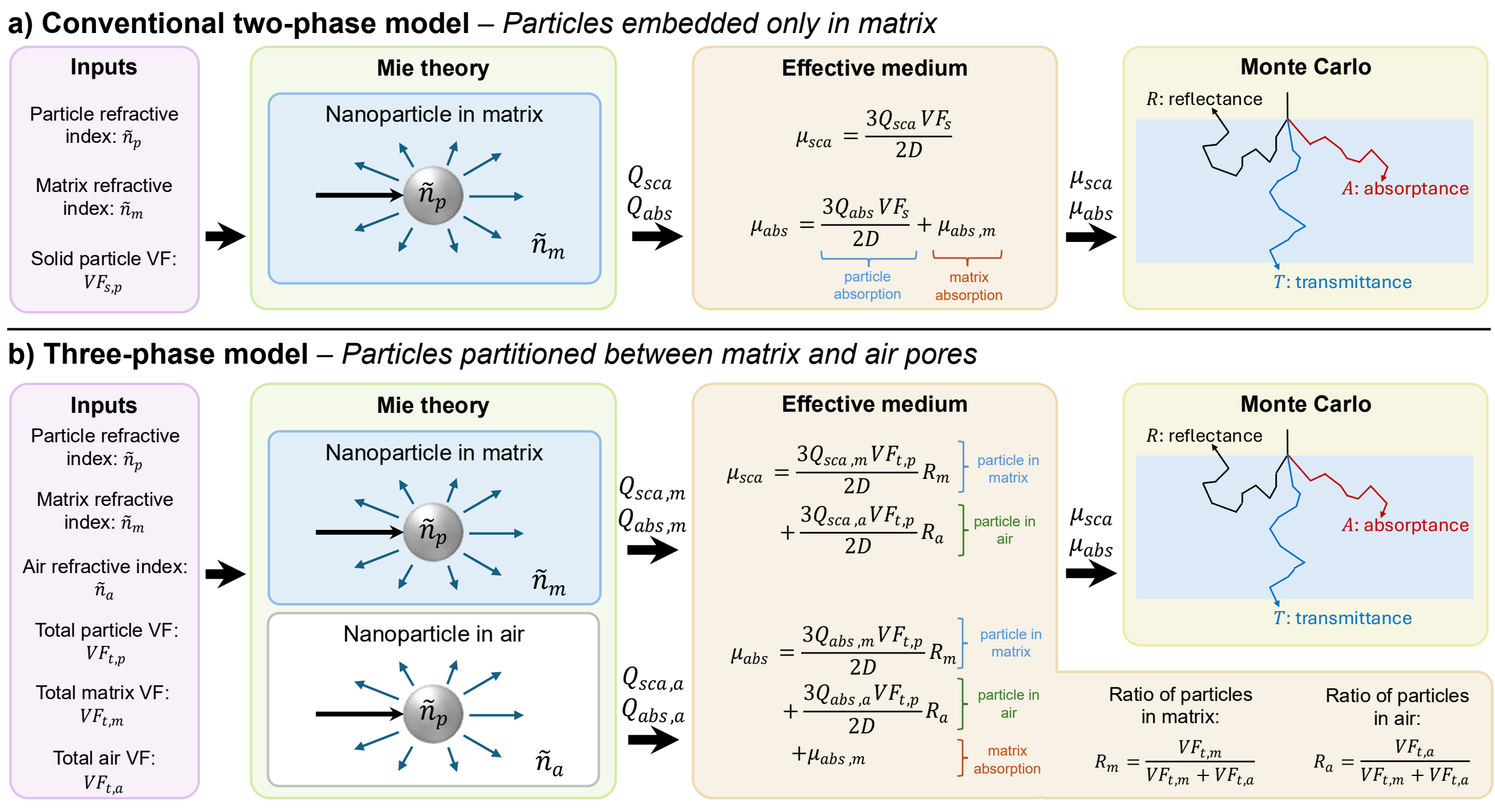

Figure 2: The flowchart illustrating the main steps for simulating the macroscopic spectral radiative properties given the pigment, binder, and air optical properties for (a) the conventional two-phase model and (b) the proposed three-phase model.

To validate the macroscopic optical properties predicted by the model and to extract experimental input parameters, the fabricated three-phase paint coatings were optically characterized across the solar spectrum (0.4 – 2.5 μm). The measured spectral reflectance and transmittance are presented in Figs. 3(a) and 3(c). For $BaSO_4$, increasing the solid volume fraction systematically elevates the reflectance and reduces the transmittance across the solar spectrum, consistent with the higher reduced scattering coefficient at greater $VF_s$. It should be noted that controlling paint thickness is challenging, as the solid volume fraction affects the wet film thickness (WFT) and its relationship to the dry film thickness (DFT). At a fixed binder-to-solvent ratio of 1:4, the calculated porosity increases approximately linearly with solid volume fraction, ranging from 27.2% to 43.8% for $BaSO_4$, given measurement uncertainties of ±20 μm in thickness and ±0.2 cm in cross-sectional side length, as shown in Fig. 3(b) (porosity calculation shown in supplementary note 4).

For $TiO_2$, the reflectance is expected to increase with $VF_s$; however, at 60% $VF_s$ the transmittance approaches zero across the visible region, which compromises the back-fitting procedure used to extract the attenuation coefficients as the spectral transmittance values in visible exceeds the saturation point. Thinner film samples were therefore used for this formulation. Furthermore, the binder-to-solvent ratio at the 60% $VF_s$ was increased to 1:5 to reduce the viscosity of the highly concentrated suspension to a level suitable for uniform paint film deposition. This increase in solvent content appears to elevate the porosity beyond the linear trend established by the 20% and 40% $VF_s$ samples, reaching approximately 62% porosity as shown in Fig. 3(d), likely due to the greater volume of solvent available to generate more voids upon evaporation. Each coating's pigment material, thickness, solid volume fractions, and total volume fractions ($VF_t$) are listed in Table 1.

Table 1: The volume fractions of $BaSO_4$ and $TiO_2$ paints in the wet state (pigment and binder) and dry state (pigment, binder, and air) with the corresponding thicknesses at each solid volume fraction level.

| | | | $VF_s$ - solid VF (%) | | $VF_t$ - total VF (%) | | |
|---|---|---|---|---|---|---|---|
| # | Pigment | Thickness (μm) | Pigment | Binder | Pigment | Binder | Air pore |
| 1 | $TiO_2$ | 175 | 20.0 | 80.0 | 14.3 | 57.3 | 28.4 |
| 2 | $TiO_2$ | 228 | 40.0 | 60.0 | 25.1 | 37.6 | 37.3 |
| 3 | $TiO_2$ | 110 | 60.0 | 40.0 | 22.5 | 15.0 | 62.5 |
| 4 | $BaSO_4$ | 120 | 20.0 | 80.0 | 14.6 | 58.2 | 27.2 |
| 5 | $BaSO_4$ | 255 | 40.0 | 60.0 | 26.1 | 39.1 | 34.8 |
| 6 | $BaSO_4$ | 320 | 60.0 | 40.0 | 33.7 | 22.5 | 43.8 |

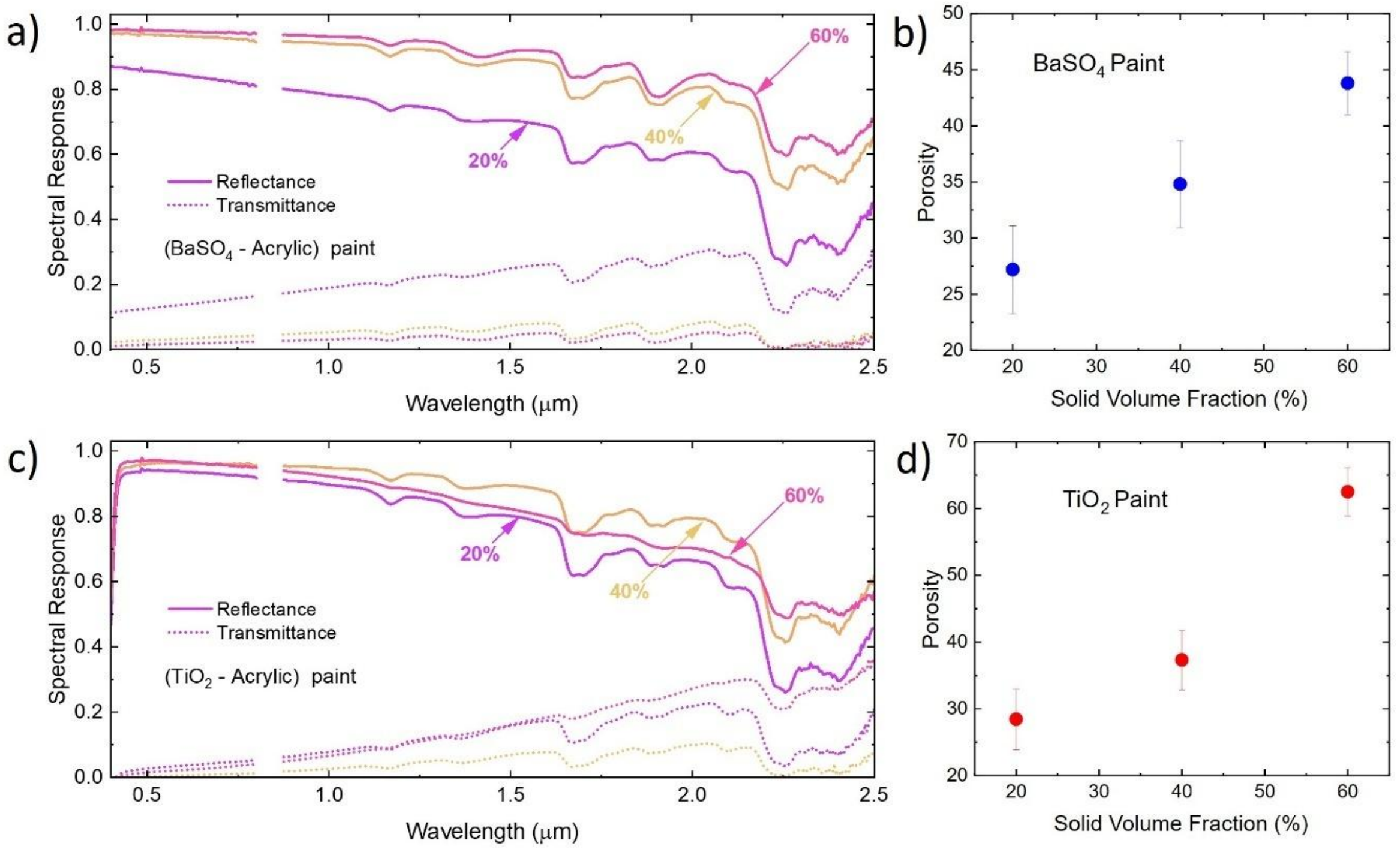

Figure 3: The measured spectral reflectance and transmittance using a UV-Vis device and the corresponding porosity with propagated uncertainty for $BaSO_4$ paint (a, b) and $TiO_2$ paint (c, d).

Although the spectral reflectance and transmittance are the experimentally measured quantities, they are not always the most informative basis for comparison between simulation and experiment for two reasons. First, reflectance and transmittance can become insensitive as they approach 0% or 100%, such that a small difference in the measured spectral response may correspond to a very large difference in scattering properties. Second, in spectral regions where the coating extinction coefficient is low, absorption can be strongly influenced by impurities, defects, and processing-dependent material quality. This absorption affects both reflectance and transmittance, making it difficult to isolate the role of scattering from the measured spectra alone. For these reasons, the reduced scattering coefficient and absorption coefficient are recovered through an inverse Monte Carlo method, assuming a planar coating with homogeneous optical properties. Direct comparison between the predicted and recovered reduced scattering coefficients provides a more sensitive and physically meaningful basis for evaluation light scattering within the coating and assessing model performance. Figure 4 shows the recovered properties for the three $BaSO_4$ coatings in Figs. 4(a-b) and for the three $TiO_2$ coatings in Figs. 4(c-d).

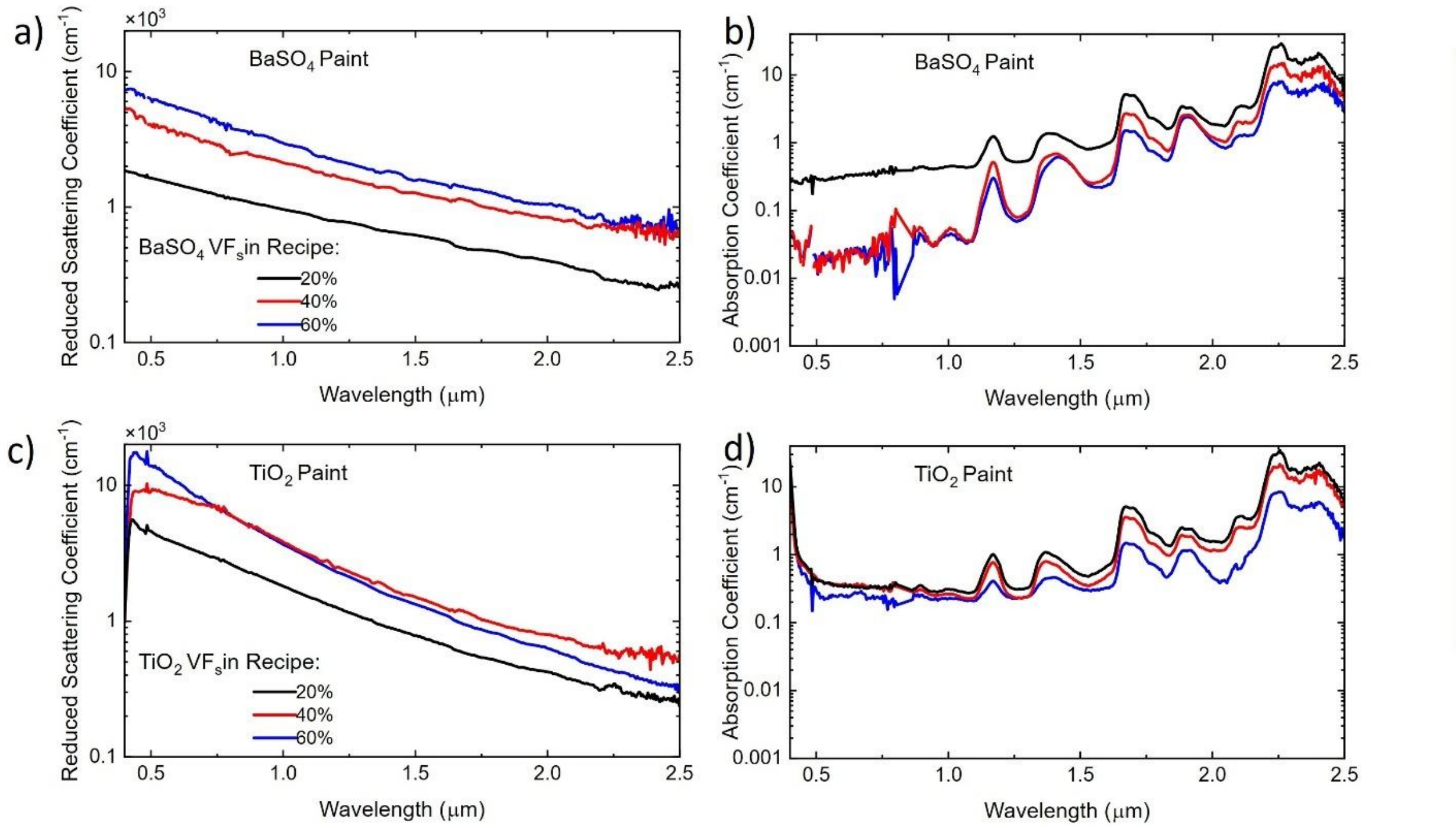

Figure 4: The spectral backfitted reduced scattering coefficient and absorption coefficient for $BaSO_4$ paint (a, b) and $TiO_2$ paint (c, d) at varied solid volume fractions shown on a logarithmic scale.

Figure 5 compares the reduced scattering coefficients obtained from the experimental backfit, two-phase model, two-phase model with correction, and three-phase model for all six coatings. First consider the two-phase model versus the experimental backfit data. For the $BaSO_4$ coatings at 20%, 40%, and 60% solid nanoparticle volume fractions (Fig. 5(a)), the experimental reduced scattering coefficients are substantially higher than the two-phase model predictions, by a factor of 6.89× on average. In contrast, the opposite trend is observed for the $TiO_2$ coatings (Fig. 5(b)), where the experimental reduced scattering coefficients are lower than the two-phase model predictions, by a factor of 3.49× on average. As these coatings are at the same solid volume fractions and similar total volume fractions, this discrepancy is unlikely to be explained by dependent scattering effects alone. These results highlight the limitations of a two-phase model, which cannot capture the contrasting trends observed in $BaSO_4$ and $TiO_2$ coatings. Particularly, this model does not explain why porous $BaSO_4$ coatings exhibit substantially higher scattering and reflectance than predicted. Prior studies on $BaSO_4$ and $CaCO_3$ coatings empirically applied an augmented correction to increase the predicted scattering coefficient and partially account for this underprediction seen relative to experiment [1], [2]. As shown in Fig. 5(a), this correction brings the predicted reduced scattering coefficient closer to the experimental values for $BaSO_4$, but scattering remains underpredicted. For the $TiO_2$ coatings in Fig. 5(b), however, this correction moves the reduced scattering coefficient further away from the experimental values.

The three-phase model for the $BaSO_4$ coating in Fig. 5(a) substantially increases the reduced scattering coefficient over the two-phase model prediction. This increase occurs because $BaSO_4$ has a relatively small refractive index contrast with the acrylic binder, but a much larger refractive index contrast with the air pores. Although the inclusion of air pores reduces the particle volume fraction, the increased average refractive index contrast dominates, leading to a larger reduced scattering coefficient. The $TiO_2$ coatings show a different trend in Fig. 5(b), where the three-phase model predicts a reduced scattering coefficient similar to that of the two-phase model.

This occurs because $TiO_2$ already possesses a large refractive index contrast with the acrylic binder. Therefore, although the inclusion of air pores further increases the refractive index contrast, the accompanying reduction in the $TiO_2$ volume fraction largely offsets this benefit. This three-phase model provides a mechanistic explanation for the experimental observation that porous $BaSO_4$ coatings significantly outperform two-phase model predictions, whereas porous $TiO_2$ coatings do not exhibit this same enhancement. These results suggest a general design principle: the inclusion of air pores in nanoparticle composite coatings can strongly enhance scattering when the refractive index contrast between the particle and matrix is small, but provides limited additional benefit when this contrast is already large.

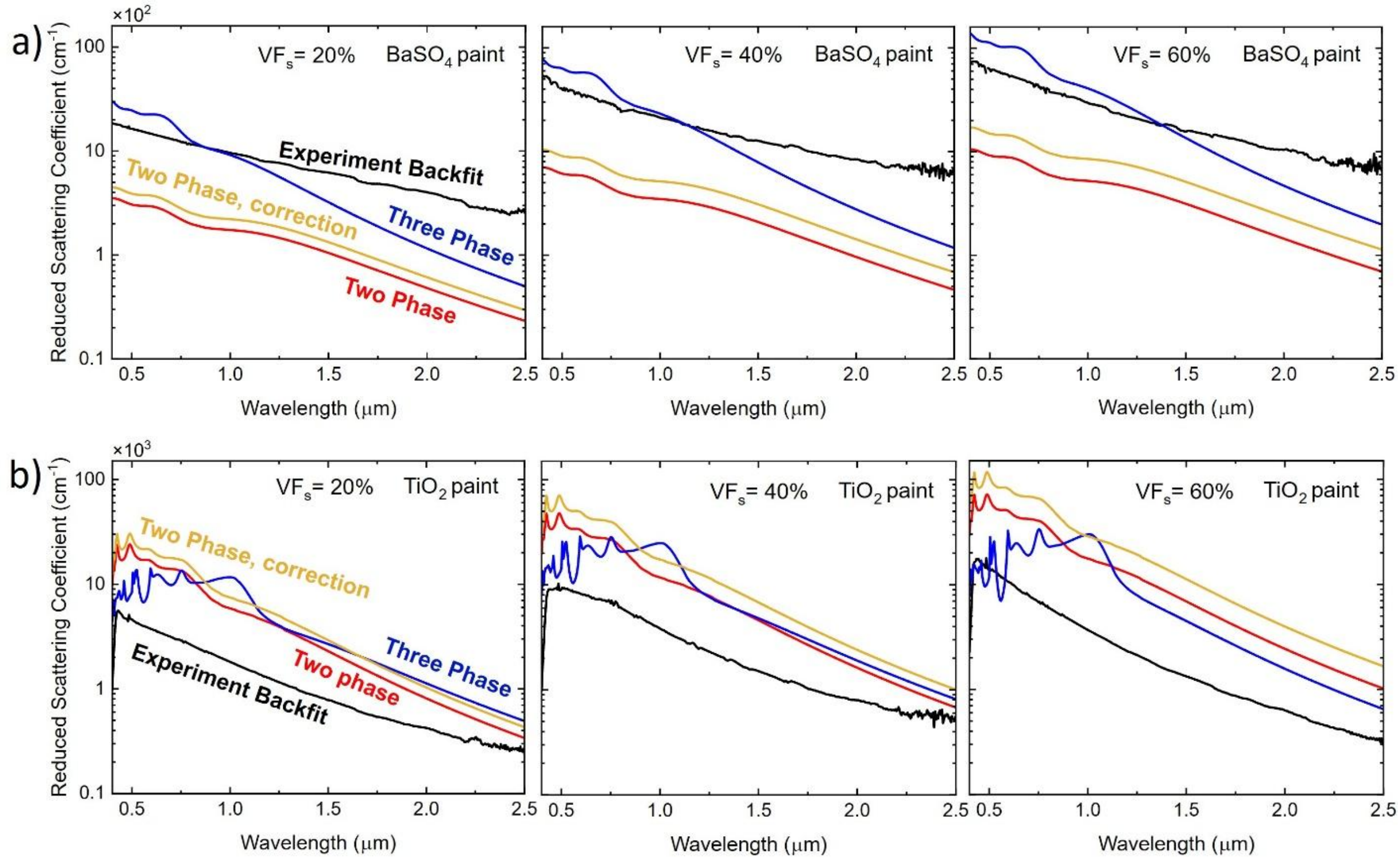

Figure 5: The experimentally backfitted reduced scattering coefficient serves as a benchmark against three computational approaches: the conventional two-phase model with and without augmented scattering correction, and the proposed three-phase model, across different solid volume fractions. (a) Reduced scattering coefficient for $BaSO_4$ paint. (b) Reduced scattering coefficient for $TiO_2$ paint.

Although the primary objective of the proposed three-phase model is to explain the experimentally observed trends, the mean absolute logarithmic error of the predicted reduced scattering coefficient relative to the experimental backfit is provided in Table 2 (metric defined in supplementary note 1). This metric quantifies the average magnitude of prediction error on a logarithmic scale. This is used because the reduced scattering coefficient covers several orders of magnitude, and errors in optical property predictions are more meaningfully interpreted in terms of relative differences. The results in Table 2 show on average, the two-phase and two-phase + correction model have similar error, with the two-phase + correction model performing better on $BaSO_4$ but worse on $TiO_2$ coatings. The three-phase model has approximately half the error (0.379 vs. 0.667) relative to the two-phase models.

Table 2: The reduced scattering coefficient prediction mean absolute log error across different solid volume fractions for $BaSO_4$ and $TiO_2$ paints comparing the conventional two-phase model with and without augmented scattering correction and the proposed three-phase model.

| Model | # 1 | # 2 | # 3 | # 4 | # 5 | # 6 | Mean |
|---|---|---|---|---|---|---|---|
| two-phase | 0.435 | 0.437 | 0.655 | 0.813 | 0.861 | 0.800 | 0.667 |
| two-phase + correction | 0.538 | 0.608 | 0.867 | 0.710 | 0.691 | 0.589 | 0.667 |
| three-phase | 0.494 | 0.452 | 0.479 | 0.315 | 0.292 | 0.241 | 0.379 |

Figure 6 compares the spectral reflectance and transmittance from the experiment, two-phase model, and three-phase model for all six coatings. For the $BaSO_4$ coatings in Fig. 6(a), the three-phase model provides improved predictions of both reflectance and transmittance relative to the two-phase model. For the $TiO_2$ coatings in Fig. 6(b), although the reduced scattering coefficient predicted by the two-phase and three-phase models are similar, the three-phase model overpredicts reflectance relative to the two-phase model. This occurs because the three-phase model has a lower spectral absorption coefficient due to the reduced volume fraction of the weakly absorbing acrylic matrix. However, sub-band gap absorption is commonly observed in $TiO_2$ due to intrinsic defects such as oxygen vacancy and high surface-state activity, and can be further enhanced by dependent absorption at high particle loading [7], [14]. This additional absorption, which is not captured by the extinction coefficient used in the present models, could explain the lower experimental reflectance relative to the model predictions. Therefore, direct comparison of the reduced scattering coefficient provides a more robust metric for evaluating light scattering models, because it is less confounded by uncertainty in the absorption coefficient [15] (absorption coefficient comparison across models in supplementary note 2).

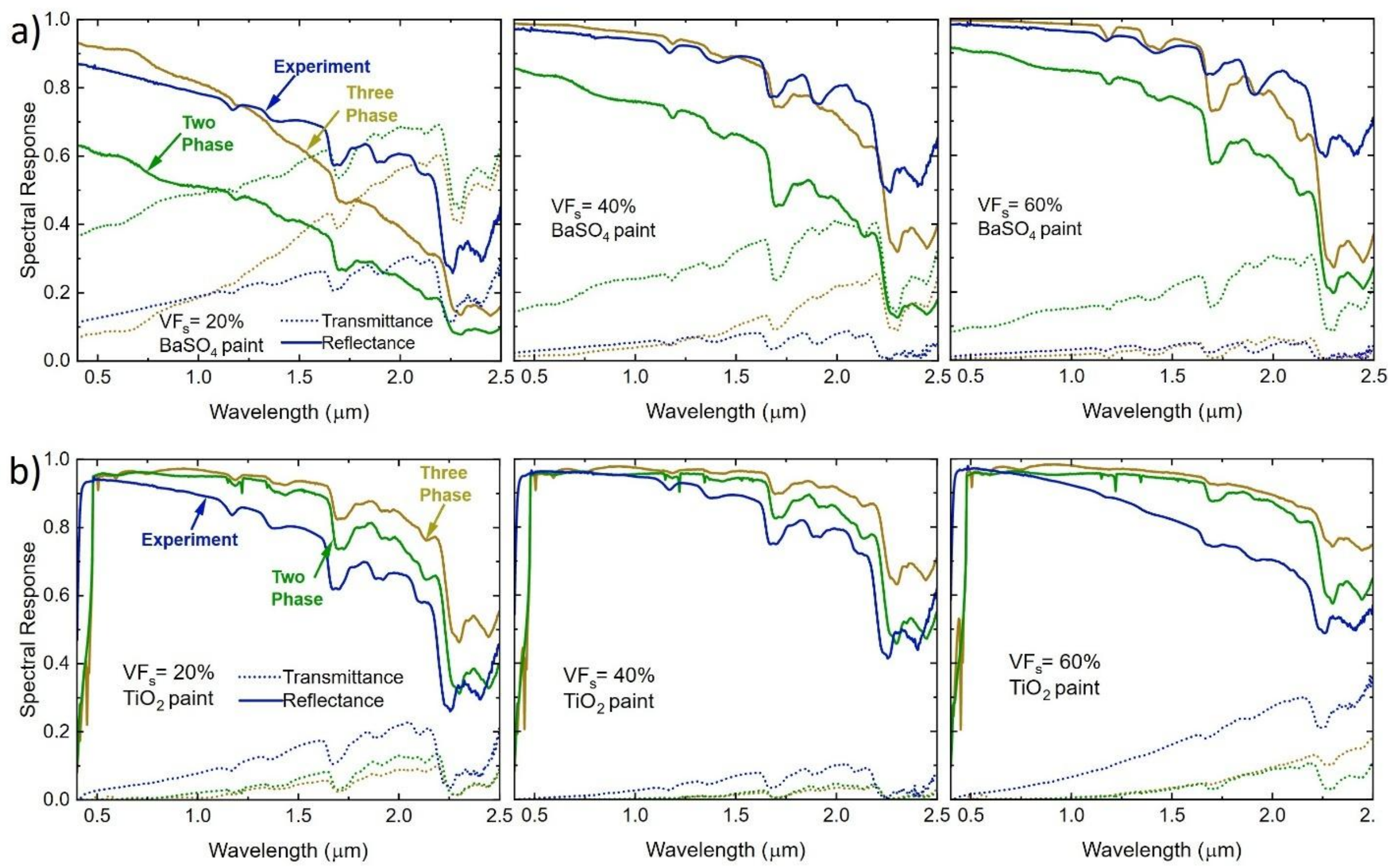

Figure 6: The spectral reflectance and transmittance simulated using the conventional two-pphase model and the proposed three-phase model in comparison to experiment across different solid volume fractions for (a) $BaSO_4$ paint and (b) $TiO_2$ paint.

While the proposed three-phase optical model demonstrates promising agreement with experimental measurements and establishes a foundation for optical modeling in three-phase coating systems, it is important to acknowledge the scope of its applicability and the assumptions. One area requiring further investigation concerns the influence of pigment particle size and morphology on the resulting air pore volume fraction. The present study does not systematically address this dependence, in part because commercially available fillers with distinct particle sizes spanning the near-infrared and upper visible ranges are limited, and because synthesis impurities in vendor-supplied materials introduce uncertainties that are difficult to decouple from size-dependent optical effects.

A further limitation is the treatment of interparticle interactions. The present model adopts the independent scattering assumption, under which each particle is excited solely by the incident wave. At the solid volume fractions considered here, however, each particle is in practice also excited by the scattered fields of neighboring particles — an effect known as dependent scattering. While the independent scattering approximation yields promising agreement with the experimental back-fitting results, as shown in Table 2, improved accuracy is expected once dependent scattering is incorporated into the three-phase model. The nature of this correction differs between the $BaSO_4$ and $TiO_2$ paints (with small and large relative refractive indices) and across the solar spectrum (with medium absorption in near infrared region). Dependent scattering is a complex phenomenon, where its effect on scattering cross sections has yielded conflicting results in the literature, with studies reporting both increases and decreases for a randomly distributed particle ensemble when the clearance-to-wavelength ratio decreases [7], [8], [9]. This contradiction has been addressed in a recent study proposing better validated prediction and a criterion for the onset of dependent scattering, by excluding the coherent scattering and extracting only the incoherent part of the ensemble [15]. However, that study is limited to non-absorbing media, which is violated in the near-infrared region where the acrylic matrix is absorbing. In absorptive media, dependent scattering modifies the local field experienced by each particle and alters the partitioning between scattering and absorption at the single-particle level in a manner not captured by independent scattering models [8], [16], [17]. Furthermore, treating dependent scattering simultaneously for dense particles embedded in both matrix and air phases is particularly challenging. Future extensions of the three-phase model could directly address this limitation and improve model fidelity at higher pigment concentrations, either by incorporating the static structure factor to capture far field interference or by using full multiple-scattering T-matrix calculations to capture both near field and far field effects.

The experimental results and proposed three-phase model introduced here are intended to serve as a benchmark for future work in the development of three-phase nanocomposite medium scattering theory and simulation methodology. To encourage this work, the material properties used in the proposed model, the experimentally recovered and model predicted spectral reduced scattering coefficient, and the experimental and model predicted spectral reflectance and transmittance are open sourced through the Harvard Dataverse dataset entitled “Three-Phase Nanocomposite Optical Data” at https://doi.org/10.7910/DVN/VJEWXE.

## CONCLUSION

In summary, this work experimentally characterized photon scattering in porous three-phase nanoparticle composite coatings and proposed a three-phase model to account for the nanoparticles, matrix, and air pores observed in high particle volume fraction coatings. Six $BaSO_4$-acrylic and $TiO_2$-acrylic coatings were fabricated at varying particle volume fractions and optically characterized from 0.4 – 2.5 μm wavelength. The experimental reflectance and transmittance were used to recover the reduced scattering coefficient and absorption coefficient through an inverse Monte Carlo method, enabling direct comparison between the measured photon scattering and model predictions. The conventional two-phase model substantially underpredicts scattering in the $BaSO_4$ coatings while overpredicting scattering in the $TiO_2$ coatings. This contrasting behavior demonstrates that porous nanoparticle coatings cannot be accurately described by two-phase models. The proposed three-phase model provides a physical explanation for the observed experimental trends. For $BaSO_4$ coatings, the inclusion of air pores introduces a significantly larger refractive index contrast than the original $BaSO_4$-acrylic interface provides, which substantially increases scattering over the two-phase model prediction. However, for the $TiO_2$ coatings, the $TiO_2$-acrylic refractive index contrast is already large, so the additional contrast introduced by the air pores is largely offset by the reduced total particle volume fraction occurring when air pores form. Across all six coatings, the mean absolute log error for reduced scattering coefficient prediction for the two-phase and three-phase models are 0.667 and 0.379, respectively. These results suggest a general design principle: the inclusion of air pores in nanoparticle composite coatings can strongly enhance scattering when the refractive index contrast between the particle and matrix is small, but provides limited additional benefit when this contrast is already large. While the proposed model is simple and does not capture three-phase dependent scattering effects or complex air pore morphology, it does explain the observed experimental trends and provides a useful benchmark for future three-phase model development. The experimental measurements and model predictions presented here are intended to motivate the development of more rigorous three-phase nanocomposite medium scattering theory and simulation methodology.

# METHADOLOGY

## *Experimental Methodology*

Paint samples were prepared using poly(methyl methacrylate) (PMMA) as the matrix, selected for its high optical transparency across the solar spectrum, with two white filler materials of distinct refractive indices. For each formulation, the filler was first dispersed in solvent and probe-sonicated for 15 minutes at 20% amplitude with a 20 s on / 40 s off pulsing cycle to break up agglomerates and achieve a uniform dispersion. The suspension was then degassed through repeated cycles to remove air bubbles, and allowed to cool under stirring at 340 rpm before matrix addition as PMMA is sensitive to elevated temperatures. The PMMA powder was subsequently introduced slowly into the cooled suspension under continuous stirring to ensure complete dissolution. Tripropylene glycol monomethyl ether (TPM) was added as a coalescing agent at 0.3 wt% of the combined solution mass to promote film formation. Three solid volume fractions were prepared for each filler — 20%, 40%, and 60% — with the rest going to the acrylic matrix portion. The binder-to-solvent mass ratio was fixed at 1:4 for all formulations except the 60% $TiO_2$ samples, for which it was adjusted to 1:5 to achieve a workable viscosity. All formulations were left stirring for a day prior to deposition. Paint films were deposited onto glass slides using a metal applicator blade to control thickness. The coated slides were dried in a fume hood under ambient airflow for 5 to 7 days to ensure complete solvent evaporation before optical and structural characterization.

To estimate the air porosity of the dried paint films, both the total geometric volume and the solid material volume were determined experimentally. After drying for at least one week, the films were carefully detached from the glass substrate and square section from the middle of sample were cut for mass and cross-sectional measurements. The sections were then mounted on clean glass slides using tape for use during optical characterization. Glass substrate was selected as the substrate for its high solar transparency, with the caveat that some residual absorption may be present below approximately 320 nm in the near-ultraviolet region.

*Existing two-phase model*

There are various methods for simulating the spectral response of two-phase nanoparticle-matrix composites, from direct simulation of Maxwell's equations to hybrid models which solve for the medium's optical properties before solving the radiative transfer equation. The method selection often depends upon the coating thickness and the scattering regime, which is a function of the size parameter $x = \pi d/\lambda$ where $d$ is the nanoparticle diameter and $\lambda$ is the photon wavelength. As the coatings investigated here cover a broad wavelength range encompassing all three scattering regimes, including wavelengths from 0.4 – 2.5 μm with 0.4 μm $BaSO_4$ and 0.3 μm $TiO_2$ nanoparticles, a three-step model shown in Fig. 2(a) is implemented which is commonly applied to model nanoparticulate coatings [18], [19], [20]. First, Mie theory calculates the absorption efficiency ($Q_{abs}$), scattering efficiency ($Q_{sca}$), and asymmetry parameter ($g$) of individual nanoparticles within the matrix. Second, effective medium theory volume averages these properties to calculate the scattering coefficient ($\mu_{sca}$) and absorption coefficient ($\mu_{abs}$) of an effective homogenous medium. Third, Monte Carlo simulations stochastically model photon transport through the effective medium to determine the spectral response, including the reflectance, absorptance, and transmittance. The open-source program FOS is used to implement this method [13].

This two-phase method assumes independent scattering. At high particle volume fractions, dependent scattering occurs where the scattering properties of individual particles is affected by near-field and far-field interactions of light scattering from nearby particles [21], [22]. Typically, at the particle size parameter and volume fraction studied, the dependent scattering coefficient ($\mu_{sca,dep}$) is less than the independent scattering coefficient ($\mu_{sca}$). However, spectral measurements from radiative cooling paint literature studying $CaCO_3$ [2] and $BaSO_4$ [1] coatings show higher reflectance and reduced transmittance compared to two-phase independent scattering simulations, indicating an increased scattering coefficient. To account for this, prior works [1], [2], [10], [23] applied an augmented scattering correction based on the solid volume fraction of particles ($VF_s$) formulated by Kamiuto [24] which was originally intended for large particle sizes:

$$\mu_{sca,dep} = \mu_{sca}[1 + (3/2)VF_s - (3/4)VF_s^2]. \quad (1)$$

These studies empirically apply this correction to account for the combined effects of both dependent scattering and porosity to bring simulation results closer to experimental measurements. This study applies this correction when specifically listed as the two-phase model with correction.

*Three-phase model*

The three-phase model implemented in this work, as shown in Fig. 2(b), builds upon the foundation of the two-phase framework. three-phase coatings are complex mixtures of nanoparticles, matrix, and air pores. The prior two-phase model only accounted for the solid volume fractions ($VF_s$) within the paint. To account for three phases, the total volume fraction of particles ($VF_{t,p}$), matrix ($VF_{t,m}$), and air pores ($VF_{t,a}$) must be considered. This work assumes the nanoparticles are embedded either in air pore or in matrix, and the ratio of each is based on the volume fractions. The ratio of nanoparticles assumed to be embedded in matrix is defined as:

$$R_m = \frac{VF_{t,m}}{VF_{t,m} + VF_{t,a}} \tag{2}$$

and the ratio of nanoparticles assumed to be embedded in air pores is defined as:

$$R_a = \frac{VF_{t,a}}{VF_{t,m} + VF_{t,a}}. \tag{3}$$

The Mie theory framework by Frisvad [12] for absorbing mediums is implemented to calculate the scattering efficiency, absorption efficiency, and asymmetry parameter separately for particles embedded in air ($Q_{sca,a}, Q_{abs,a}, g_a$) and particles embedded in matrix ($Q_{sca,m}, Q_{abs,m}, g_m$), respectively. If there are multiple particle sizes or materials, this process is repeated for each unique particle. Then, a novel effective medium approach is implemented where these properties are volume averaged based on the percent of particle assumed to be embedded within air and matrix to convert these properties into a bulk homogeneous medium. The scattering coefficient is calculated by:

$$\mu_{sca} = \sum_{i=1}^{P} \frac{3Q_{sca,i,m}VF_{t,p}}{2D_i} R_m + \sum_{i=1}^{P} \frac{3Q_{sca,i,a}VF_{t,p}}{2D_i} R_a \tag{4}$$

where the $i$ subscript denotes individual particle size and material for $P$ number of particles. The absorption coefficient is calculated by:

$$\mu_{abs} = \sum_{i=1}^{P} \frac{3Q_{abs,i,m}VF_{t,p}}{2D_i} R_m + \sum_{i=1}^{P} \frac{3Q_{abs,i,a}VF_{t,p}}{2D_i} R_a + \frac{4\pi\kappa_m VF_{t,m}}{\lambda} \tag{5}$$

where $\kappa_m$ is the extinction coefficient of the matrix to account for matrix absorption. The asymmetry parameter is calculated by:

$$g = \frac{1}{\mu_{sca}} \left( \sum_{i=1}^{P} \frac{3Q_{sca,i,m}VF_{t,p}g_{i,m}}{2D_i} R_m + \sum_{i=1}^{P} \frac{3Q_{sca,i,a}VF_{t,p}g_{i,a}}{2D_i} R_a \right). \tag{6}$$

This approach can be considered as a post-Mie theory effective medium model. Alternatively, the refractive index between the matrix and air could be averaged in a pre-Mie theory effective medium model. However, this approach would fail when the particle refractive index is between the air and matrix refractive index. Finally, the Monte Carlo method implemented in FOS [13] models photon transport within the effective bulk medium to calculate the spectral response.

Due to the complexities of three-phase dependent scattering, this model assumes independent scattering. The main aim of this model is to begin understanding the effects of air pores embedded in nanoparticle composites to explain trends seen in experiments. The authors encourage future studies to consider dependent scattering in multi-phase systems.

*Optical Property Backfitting*

The forward Monte Carlo method simulates the reflectance and transmittance as a function of the scattering coefficient, absorption coefficient, asymmetry parameter, refractive index, and coating thickness. This forward process is straightforward to solve. The inverse Monte Carlo method aims to solve for optical properties as a function of the reflectance and transmittance. This is complex to directly simulate, and is often solved by optimizing the optical property inputs into the forward Monte Carlo method until the reflectance and transmittance match [25]. Here, an inverse Monte Carlo method is applied to determine the reduced scattering coefficient ($\mu_s' = \mu_s(1-g)$) and the absorption coefficient of the experimentally demonstrated coatings. This is formulated as an optimization problem with the aim of minimizing the mean squared error ($E$) between the simulated spectral reflectance ($R_{sim}$) and transmittance ($T_{sim}$) and the experimental spectral reflectance ($R_{exp}$) and transmittance ($T_{exp}$):

$$\underset{\mu_{a,j},\mu_{s,j}'}{minimize} \sum_{j=1}^{N} \left[ \left( R_{sim}(\lambda_j, \mu_{a,j}, \mu_{s,j}') - R_{exp}(\lambda_j) \right)^2 + \left( T_{sim}(\lambda_j, \mu_{a,j}, \mu_{s,j}') - T_{exp}(\lambda_j) \right)^2 \right] \tag{7}$$

$$\text{subject to} \quad \mu_{a,j} \geq 0 \tag{8}$$
$$\mu_{s,j}' \geq 0$$

where $j$ is the wavelength index across $N$ number of wavelengths. The error is minimized by iteratively updating the absorption and reduced scattering coefficients through a gradient descent method:

$$\mu_{a,j,t+1} = \mu_{a,j,t} - \alpha \frac{\partial E(\lambda_j, \mu_{a,j}, \mu_{s,j}')}{\partial \mu_{a,j,t}} \tag{9}$$

$$\mu_{s,j,t+1}' = \mu_{s,j,t}' - \alpha \frac{\partial E(\lambda_j, \mu_{a,j}, \mu_{s,j}')}{\partial \mu_{s,j,t}'} \tag{10}$$

where $t$ is the gradient descent iteration, $\alpha$ is the learning rate, and $E$ is the mean squared error. The gradients are estimated numerically by applying a small perturbation to the optical properties in the forward Monte Carlo method. This inverse Monte Carlo method takes the experimental spectral reflectance and transmittance, and calculates the reduced scattering coefficient and absorption coefficient for direct comparison to the proposed model. This method assumes the medium is planar and homogenous.

**Data Availability Statement**

The original results of the study are available from the corresponding authors upon reasonable request.

**Supporting Information**.

The Supporting Information is available free of charge.

Additional modeling details are provided through the mean absolute log error. The discussion on absorption coefficients across the two-phase, two-phase with correction, and three-phase models. More details on porosity of nanocomposite paint, the porosity calculation, and SEM images and details on the experimental parameters.

**AUTHOR INFORMATION**

**Corresponding Authors**

**Xiulin Ruan** − School of Mechanical Engineering and The Birck Nanotechnology Center, Purdue University, West Lafayette, Indiana 47907-2088, United States; orcid.org/ 0000-0001-7611-7449; Email: ruan@purdue.edu

**Daniel Carne** − Department of Mechanical and Nuclear Engineering, United States Naval Academy, Annapolis, MD 21402, United States; orcid.org/ 0009-0009-1531-8189

**Authors**

**Khalid Alhammadi** − School of Mechanical Engineering and The Birck Nanotechnology Center, Purdue University, West Lafayette, Indiana 47907-2088, United States; orcid.org/0009-0008-6176-3323; Email: alhammak@purdue.edu. Mechanical Engineering Department, King Saud University, P.O. Box 800, Riyadh, 11421, Saudi Arabia

**Author Contributions**

X.R., D.C. and K.A. conceived and designed the study. K.A. and D.C. implemented the model. K.A. performed the experimental fabrication and measurement, analyzed the results, and wrote the manuscript. D.C. applied the back-fitting models and error calculation and wrote the manuscript.

X.R. supervised the project and edited the paper. All authors contributed to discussions and revisions of the manuscript.

**Disclaimer**

The views expressed in this article are those of the author(s) and do not reflect the official policy or position of the U.S. Naval Academy, Department of the Navy, the Department of War, or the U.S. Government.

## ACKNOWLEDGMENT

K.A. acknowledges the support provided by the King Saud University scholarship . X.R. acknowledges the partial support from National Science Foundation through the award # 2621107. During the preparation of this work the author used Claude Opus 4.8 for only proofreading and editorial refinements during the manuscript preparation. Extreme care was taken to review and verify all AI-assisted edits to ensure scientific accuracy and clarity. After using this tool, the authors reviewed and edited the content as needed and take full responsibility for the content of the published article.

# Supporting Information:

# A three-phase framework for photon scattering in porous nanocomposites

Khalid Alhammadi[a,b], Daniel Carne[c,*], Xiulin Ruan[a,*]

[a] School of Mechanical Engineering, Purdue University, West Lafayette, IN 47907, USA
[b] Mechanical Engineering Department, King Saud University, P.O. Box 800, Riyadh, 11421, Saudi Arabia
[c] Department of Mechanical and Nuclear Engineering, United States Naval Academy, Annapolis, MD 21402, USA
[*] Corresponding Authors: ruan@purdue.edu, carne@usna.edu

## Supplementary Notes

### Supplementary Note 1: Mean absolute log error

As the change in spectral response exponentially decays at increasing scattering coefficients, we report reduced scattering coefficient error as the mean absolute log error ($MALE$). The resultant error value represents how many orders of magnitude the prediction is off relative to the experimental result. This is defined as:

$$MALE = \frac{1}{N}\sum_{j=1}^{N}\left|\frac{log_{10}(\mu'_{sca,model})}{log_{10}(\mu'_{sca,experiment})}\right| \quad (1)$$

### Supplementary Note 2: Absorption coefficients for conventional and proposed models.

The absorption coefficient comparisons for both paints at different solid volume fractions are shown in Supplementary Fig. 1(a,b). The data are back-fitted from experimental measurements and estimated using the conventional two-phase model, the two-phase model with correction, and the three-phase model. The three-phase model shows good agreement with the experimental data.

Interestingly, in the near infrared region where medium absorption exists, the correction drives the two-phase model further from the experimental data for both paints, worsening its overestimation. The absorption coefficients of both paints in near infrared region follow the same trend and are of roughly the same order of magnitude for the three-phase model and the back-fitted data. The presence of a low-volume-fraction filler in an absorbing medium leads to a higher absorption coefficient, resulting from the local absorption enhancement by a scattering particle in an absorbing medium. This effect has been confirmed by Monte Carlo simulations, where varying the filler volume fraction and thus the scattering coefficients produced a Gaussian-like overall absorption coefficient due to the scattering particle-to-medium absorption. At high solid pigment volume fractions, however, this absorption coefficient becomes much lower, where most radiative cooling paints exist.

## Supplementary Note 3: Porosity of nanocomposite paint.

The porosity of paint can be affected by many factors, not only the solid volume fraction, as discussed in literature [1], [2]. The binder–pigment interaction, the solvent evaporation rate, the substrate type, and the binder–solvent ratio are only a few to mention. For the latter, the phase change of the binder upon drying contributes substantially to air-pore formation: it depends on the amount of solvent present and whether it lies within the range that allows the binder to deform and coalesce into a continuous dried film [3]. The drying dynamics also matter, as the rate at which the solvent evaporates sets how much time the binder has to flow and close voids before it solidifies.

## Supplementary Note 4: Porosity calculation.

The porosity is determined by measuring the total film volume and the total solid volume. It is calculated as [4]:

$$\emptyset = \left(1 - \frac{Solid\ Sample\ Volume}{Total\ Sample\ Volume}\right) * 100$$

A uniform thickness and regular square shape of the sample improve the accuracy of the calculated porosity. The solid volume is estimated from the mass and density of both the pigment and binder components.

## Supplementary Note 5: SEM imaging for both porous coating.

To gain insight into the nanoscale structure of the $TiO_2$ and $BaSO_4$ porous nanocomposite media, SEM imaging was conducted using a field-emission-gun scanning electron microscope (FEI Teneo VolumeScope FEG-SEM). Imaging was performed on the surface of the paint samples. The detector and configuration parameters are shown in Figs. 2,3, with an accelerating voltage of 5 kV and a beam current of 0.1 nA. During image collection, the contrast, astigmatism, and focus were adjusted to obtain high-quality images despite the porous surface of the nanostructure. The SEM images for the $TiO_2$ paint are shown in Supplementary Fig. 2 at three solid volume fractions (20, 40, and 60%), and correspondingly for the $BaSO_4$ paint in Supplementary Fig. 3.

## Supplementary Figures:

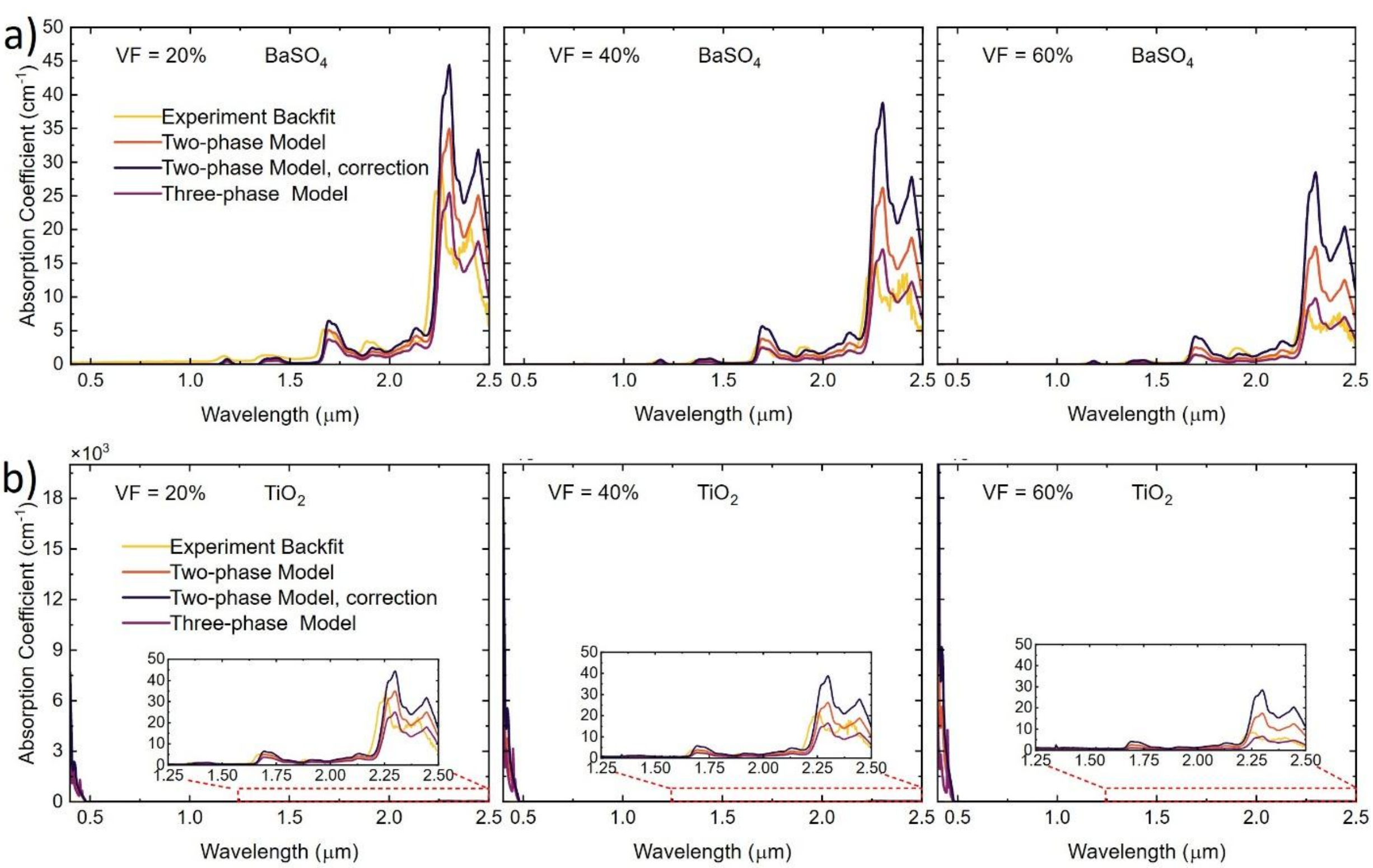

Supplementary Figure 1 (Absorption coefficient of the paint nanocomposite medium for both paints at five different solid volume fractions, calculated using the Two-phase model, the Two-phase model with correction, and the Three-phase model, and compared to the experimental back-fit. (a) $BaSO_4$ paint and (b) $TiO_2$ paint.)

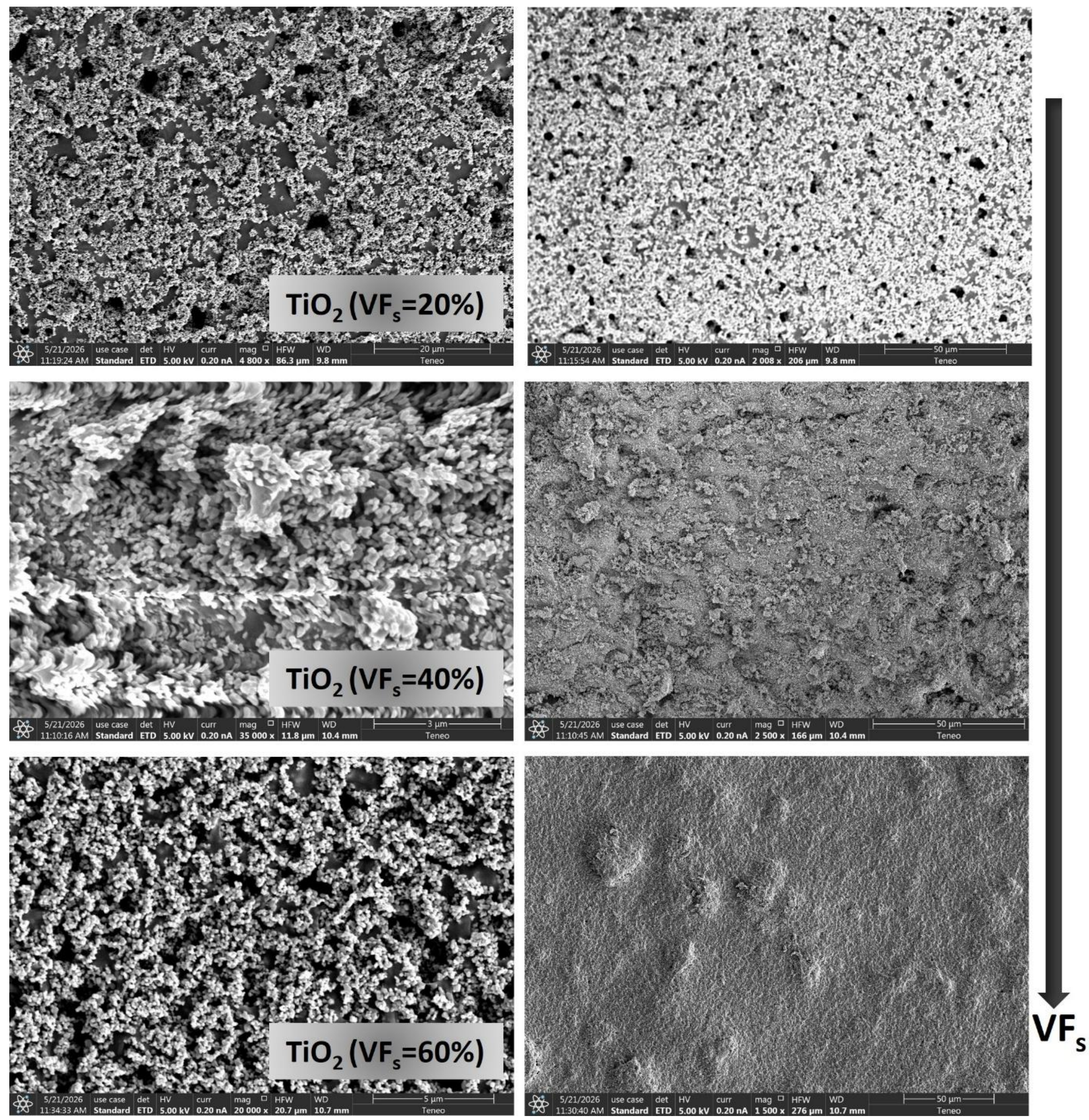

Supplementary Figure 2 (SEM images of the $TiO_2$ paint at three solid volume fractions (20, 40, and 60%), shown at multiple magnifications.)

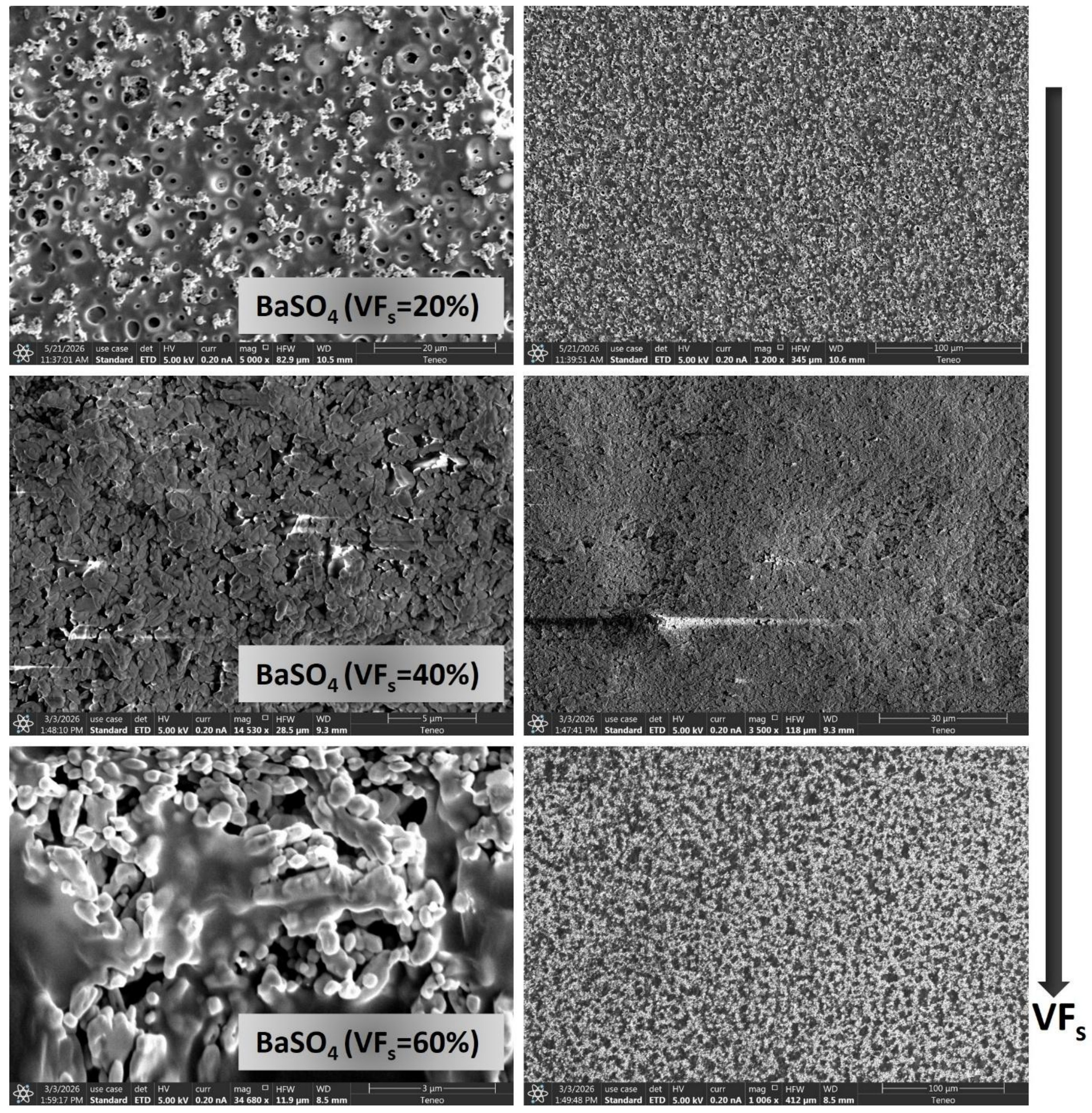

Supplementary Figure 3 (SEM images of the BaSO4 paint at three solid volume fractions (20, 40, and 60%), shown at multiple magnifications.)

## **Supplementary References**: